\documentclass[aps,prx,twocolumn, notitlepage]{revtex4-1}
\usepackage{times}
\usepackage{amsmath}
\usepackage{amsfonts}
\usepackage{amssymb}
\usepackage{dsfont}
\usepackage{graphicx}
\usepackage{physics}
\usepackage{tikz}
\usepackage{hyperref}

\usepackage[english]{babel}

\usepackage{pst-all}
\usepackage{leftidx}

\newcommand{\be}{\begin{equation}}
\newcommand{\ee}{\end{equation}}
\newcommand{\bc}{\begin{center}}
\newcommand{\ec}{\end{center}}
\newcommand{\nin}{\noindent}

\newcommand{\non}{\nonumber}

\definecolor{dualblue}{RGB}{3,101,192}

\begin{document}

\title{Universal logical gates on topologically encoded qubits via constant-depth unitary circuits}

\author{Guanyu Zhu}
\author{Ali Lavasani}
\author{Maissam Barkeshli}
\affiliation{Department of Physics, Condensed Matter Theory Center, University of Maryland, College Park, Maryland 20742, USA}
\affiliation{Joint Quantum Institute, University of Maryland, College Park, Maryland 20742, USA}

\begin{abstract}
A fundamental question in the theory of quantum computation is to understand the ultimate space-time resource costs for performing a universal set of logical quantum gates to arbitrary precision. Here we demonstrate that non-Abelian anyons in Turaev-Viro quantum error correcting codes can be moved over a distance of order the code distance, and thus braided, by a constant depth local unitary quantum circuit followed by a permutation of qubits. Our gates are protected in the sense that the length of error strings do not grow by more than a constant factor. When applied to the Fibonacci code, our results demonstrate that a universal logical gate set can be implemented on encoded qubits through a constant depth unitary quantum circuit, and without increasing the asymptotic scaling of the space overhead. These results also apply directly to braiding of topological defects in surface codes. Our results reformulate the notion of braiding in general as an effectively instantaneous process, rather than as an adiabatic, slow process.  
\end{abstract}

\maketitle

The possibility of a scalable universal quantum computer rests on quantum error correction and fault-tolerance \cite{Terhal:2015ks, Campbell:2016wb}. A promising class of quantum error correcting codes (QECCs) are topological QECCs, where the information is non-locally encoded in a topologically ordered quantum state of matter \cite{fowler2012, Terhal:2015ks, Campbell:2016wb}.
For a QECC to allow fault-tolerant quantum computation, it must be possible to perform fault-tolerant logical gates on the encoded logical qubits, for example via braiding of non-abelian anyons, holes, or twist defects in topological QECCs \cite{kitaev2003, Koenig:2010do, fowler2012, Barkeshli:T2nDmmbu2014, Cong:2016}. Braiding of Fibonacci anyons in certain non-Abelian topological QECCs \cite{levin2005,Koenig:2010do,kitaev2003} can form a universal logical gate set \cite{Freedman_Larsen_wang_2002, zhwang2010, Bonesteel:2005ho}.

Other proposed methods to realize a universal fault-tolerant gate set involve magic state distillation \cite{bravyi2005}, code
switching  or gauge fixing \cite{Paetznick:2013fu, Bombin:2015jk}. However, such methods necessarily carry a large space-time overhead depending on the code distance $d$, and all require measurements to achieve universality.
It is an open question whether a universal logical gate set can be realized with just constant-depth unitary circuits independent of code
distance $d$ and requiring no measurements.  In the context of topological codes, this question is also deeply related to a quantum complexity problem: the circuit complexity of unitary transformations between arbitrary states in the ground-state subspace of topologically ordered systems. Topologically ordered states are non-trivial phases of matter, which implies that a unitary circuit composed of few-qubit gates with depth of $\mathcal{O}(\log d)$ is needed to prepare the ground states from a trivial product state \cite{Vidal:2007va, Konig:2009bm}.  Naively one may expect that the transformation between two arbitrary ground states will also need at least a $\mathcal{O}(\log d)$-depth unitary circuit using a path connecting both states and going through the trivial product state in the middle.

Quite surprisingly, we find in this paper that braiding of non-abelian anyons, and hence universal logical gates on encoded qubits, can be performed through a constant depth unitary circuit composed of few-qubit gates acting on the physical qubits. The circuit depth is independent of the separation between the anyons, and thus independent of $d$. The braiding circuit is composed of a local quantum circuit, $\mathcal{LU}$, which implements a \textit{local geometry deformation}, and a permutation of qubits, $\mathcal{P}_\sigma: j \mapsto \sigma(j)$, separated by distance of $\mathcal{O}(d)$ and which can be implemented by a depth-2 circuit.

Due to the Solovay–Kitaev theorem \cite{nielsen_chuang_2010}, the above result also suggests that arbitrary transformations within the ground state subspace of certain topologically ordered systems supported on a punctured manifold can be realized with a constant-depth unitary circuit independent of the system size, given that the braid group representation of the topological order is computationally universal.  Furthermore, our result demonstrates, for the first time, how to construct a universal logical gate set using constant depth unitary circuits,
thus circumventing the Eastin-Knill theorem \cite{Eastin:2009cj}, where the long-range permutation is the key to circumvent the assumption in Ref.~\cite{Eastin:2009cj} (see also Ref.~\cite{Zeng:2011gs}). In addition, it also circumvents the no-go's for universality in geometrically local \cite{Bravyi:2013dx} and non-local \cite{JochymOConnor:2018is} constant-depth circuits, where the circumvention can be mainly attributed to the non-stabilizer nature of the Turaev-Viro codes.
Our result can be generalized to arbitrary braids and Dehn twists, which generate the mapping class group of genus $g$ surfaces with $n$ punctures \cite{Zhu:2017tr, Zhu:2018CodeLong} \footnote{This includes the case of transformation between arbitrary ground states supported on a closed manifold.}.

\begin{figure}[t]
\centering
\includegraphics[width=1\columnwidth]{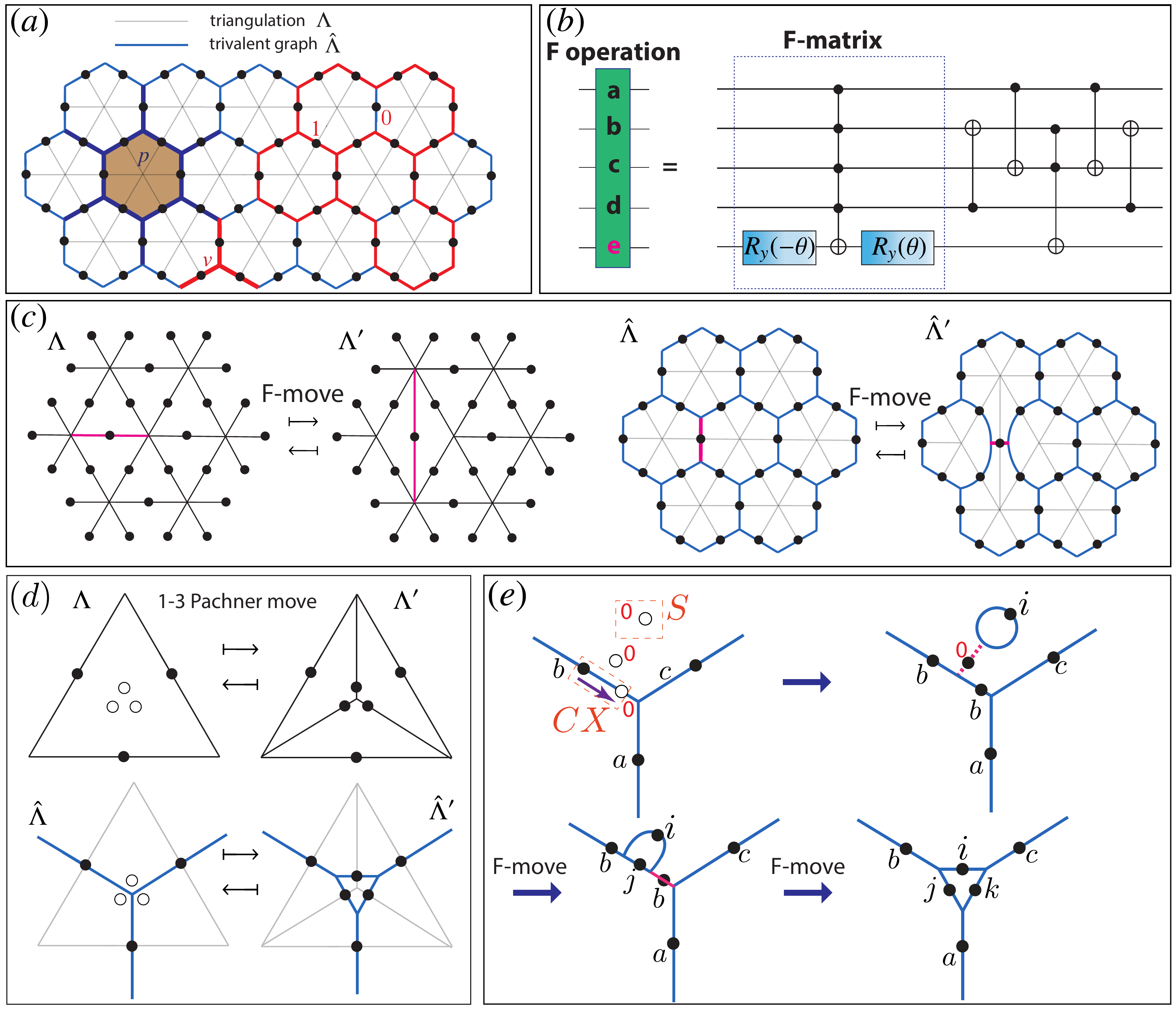}
   \caption{(a) Definition of the Levin-Wen Hamiltonian on a triangulated manifold 
   (b) Circuit for $F$ operation in Fibonacci code. (c) Definition of the 2-2 Pachner move on the triangulation $\Lambda$
and the corresponding trivalent graph defined by $\hat{\Lambda}$. The pink edges represent the edges being switched during the moves. (d,e) Definition of the 1-3 Pachner move with the addition/removal of three ancilla qubits (white dots) and the circuit implementation via unitary gates.}
\label{fig:F-move_definition}
\end{figure}

With purely local operations, a unitary circuit that moves non-Abelian anyons or defects over a distance $l$ must have a depth of $\mathcal{O}(l)$ \footnote{We note that our result here is about using unitary circuits to instantaneously move non-Abelian anyons or defects.  It is known that, without measurement errors, one can move non-abelian twist and hole defects in Abelian topological stabilizer codes instantaneously through  measurements (non-unitary) \cite{Brown:2017, fowler2012}. Nevertheless, even without measurement error, it is not known how to move non-abelian anyons instantaneously through measurements.}. Abelian anyons, in contrast, can be moved over arbitrary distances in $\mathcal{O}(1)$ time through transversal unitary operations.
Our results are thus possible because of the use of long-range permutations in one time step (between two consecutive syndrome measurements), which is naturally suitable for a variety of experimental platforms with long-range connectivity or movable qubits \cite{Linke:2017bz, Bowler:2012fr,  Wright:2013df, Home:jr, Lekitsch:2015ua, CampagneIbarcq:2017wq, Kurpiers:2017ub, Axline:2017uq, Chou:2018vz, Majer:2007em, Helmer:2009de, Naik:2017bo, HerreraMarti:2010cu}.

\textit{Turaev-viro codes.}---We consider Turaev-Viro codes \cite{levin2005, Koenig:2010do}, which can
capture all non-chiral topological states in 2D. For the application to universal gate sets we are interested in the doubled Fibonacci state realized by a specific type of Turaev-Viro code. The Turaev-Viro code associates to a closed/punctured surface $\Sigma$ a finite-dimensional code space $\mathcal{H}_\Sigma$.
We use $\Lambda$ to denote a triangulation of $\Sigma$ and $\hat{\Lambda}$ to denote the dual cellulation associated
with $\Lambda$.  More  specifically, $\hat{\Lambda}$ defines a trivalent graph, such as the
honeycomb lattice shown in Fig.~\ref{fig:F-move_definition}(a). Each edge of $\Lambda$ (equivalently, of $\hat{\Lambda}$)
is associated with an $N$-state qudit. If the qudit on a particular edge is in the state $|a\rangle$, we say that
there is a string of type $a$ passing through that edge. The wave functions in the code space can be viewed as
superpositions of closed \textit{string-net} configurations \cite{levin2005}.

The states in the code space are exact ground states of a commuting projector Hamiltonian known as the Levin-Wen Hamiltonian
\cite{levin2005},
$H_{\hat{\Lambda}}= - \sum_v Q_v -\sum_p B_p$,
where $v$ and $p$ label the vertices and plaquettes of $\hat{\Lambda}$. The 3-body vertex
projection operator $Q_v$ depends only on the three edges incident to $v$:
\begin{align}\label{branching_rules}
Q_v
\begin{tikzpicture}[baseline={([yshift=-.5ex]current  bounding  box.center)}]
\draw[thick]  (-0.6, -0.35) -- (-0.6, 0.35);
\draw[thick]   (0.5, -0.35) -- (0.6, 0);
\draw[thick]   (0.5,  0.35) -- (0.6, 0);
\draw[dualblue,thick,] (0,4/3.4/3) -- (0,0);
\draw[dualblue,thick,] (-1/3, -2/3.4/3 ) -- (0,0);
\draw[dualblue,thick] (1/3,-2/3.4/3) -- (0,0);
\draw (-1.25/3,-0.7/3.5) node {$b$};
\draw (-0.15,1.1/3.5) node {$a$};
\draw (1.25/3,-0.8/3.5) node {$c$};
\end{tikzpicture}
= \delta_{abc}
\begin{tikzpicture}[baseline={([yshift=-.5ex]current  bounding  box.center)}]
\draw[thick]  (-0.6, -0.35) -- (-0.6, 0.35);
\draw[thick]   (0.5, -0.35) -- (0.6, 0);
\draw[thick]   (0.5,  0.35) -- (0.6, 0);
\draw[dualblue,thick,] (0,4/3.4/3) -- (0,0);
\draw[dualblue,thick,] (-1/3, -2/3.4/3 ) -- (0,0);
\draw[dualblue,thick] (1/3,-2/3.4/3) -- (0,0);
\draw (-1.25/3,-0.7/3.5) node {$b$};
\draw (-0.15,1.1/3.5) node {$a$};
\draw (1.25/3,-0.8/3.5) node {$c$};
\end{tikzpicture}
\end{align}
Here, $\delta_{abc} = 0,1$ are the branching rules of the allowed string-net configuration.
The Fibonacci Turaev-Viro code has $N = 2$ and therefore each
edge of the trivalent graph contains two types of strings, as illustrated on the right side of Fig.~\ref{fig:F-move_definition}(a),
where the edges with (without) the red string correspond to an occupied (unoccupied) site
$\ket{1}$ ($\ket{0}$).  The branching rules are specified as
\be\label{branching_Fibonacci}
\delta_{abc}=\left\lbrace \begin{array}{cc}
1  & \text{if} \ abc=000,011,101,110,111, \\
0  & \text{otherwise. }
\end{array} \right.
\ee

On a honeycomb lattice [Fig.~\ref{fig:F-move_definition}(a)], $B_p$
is a 12-body operator that depends on the 6 qubits on the hexagonal plaquette and also on
the qubits on the 6 legs connecting to the hexagon. The operator can be written as
$ B_p = \sum_s d_s B_p^s/D^2$, where $d_s$ is the quantum dimension of the string label
$s$ and $D=\sum_s \sqrt{d_s^2}$ is the total quantum dimension. For the Fibonacci code, we have
$d_0=1$, and $d_1=\phi=\frac{\sqrt{5}+1}{2}$. The operator $B^s_p$ is defined via
\begin{align}
 B^s_p
\begin{tikzpicture}[baseline={([yshift=-.5ex]current  bounding  box.center)}]
\draw[thick]  (-2.3/2/1.5, -1/1.5) -- (-2.3/2/1.5, 1/1.5);
\draw[dualblue, thick, ]  (-1/2/1.5,1/4/1.5) --  (0 ,1/4/1.5+0.577/2/1.5);
\draw[dualblue, thick, ]  (1/2/1.5,1/4/1.5) --  (0 ,1/4/1.5+0.577/2/1.5);
\draw[dualblue, thick, ]  (-1/2/1.5,-1/4/1.5) --  (-1/2/1.5,1/4/1.5);
\draw[dualblue, thick, ]  (1/2/1.5,-1/4/1.5) --  (1/2/1.5, 1/4/1.5);
\draw[dualblue, thick, ]  (-1/2/1.5,-1/4/1.5) --  (0 ,-1/4/1.5-0.577/2/1.5);
\draw[dualblue, thick, ]  (1/2/1.5,-1/4/1.5) --  (0 ,-1/4/1.5-0.577/2/1.5);
\draw[dualblue, thick, ]  (-1/2/1.5,1/4/1.5) --  (-1/2/1.5-1/4/1.5 ,1/4/1.5+0.577/4/1.5);
\draw[dualblue, thick, ]  (1/2/1.5,1/4/1.5) --  (1/2/1.5+1/4/1.5 ,1/4/1.5+0.577/4/1.5);
\draw[dualblue, thick, ]  (-1/2/1.5,-1/4/1.5) --  (-1/2/1.5-1/4/1.5 ,-1/4/1.5-0.577/4/1.5);
\draw[dualblue, thick, ]  (1/2/1.5,-1/4/1.5) --  (1/2/1.5+1/4/1.5 ,-1/4/1.5-0.577/4/1.5);
\draw[dualblue, thick, ]   (0 ,1/4/1.5+0.577/2/1.5)--(0 ,1/4/1.5+0.577/2/1.5+1/4/1.5);
\draw[dualblue, thick, ]   (0 ,-1/4/1.5-0.577/2/1.5)--(0 ,-1/4/1.5-0.577/2/1.5-1/4/1.5);
\draw (-0.95/1.5,0.463/1.5) node {$a$};
\draw (-0.95/1.5,-0.463/1.5) node {$f$};
\draw (0.95/1.5,0.463/1.5) node {$c$};
\draw (0.95/1.5,-0.463/1.5) node {$d$};
\draw (0,1.05/1.5) node {$b$};
\draw (0,-0.95/1.5) node {$e$};
\draw (-0.75/2/1.5,1.066/2/1.5) node {$g$};
\draw (0.75/2/1.5,1.066/2/1.5) node {$h$};
\draw (1.3/2/1.5, 0) node {$i$};
\draw (-1.3/2/1.5, 0) node {$l$};
\draw (-0.75/2/1.5,-1.066/2/1.5) node {$k$};
\draw (0.75/2/1.5,-1.066/2/1.5) node {$j$};
\draw[thick]   (1.1/1.5, -1/1.5) -- (1.3/1.5, 0);
\draw[thick]   (1.1/1.5,  1/1.5) -- (1.3/1.5, 0);
\end{tikzpicture}
& = \sum_{g'h'i'j'k'l'} B^{s, g'h'i'j'k'l'}_{p, ghijkl,abcdef}
\begin{tikzpicture}[baseline={([yshift=-.5ex]current  bounding  box.center)}]
\draw[thick]  (-2.3/2/1.5, -1/1.5) -- (-2.3/2/1.5, 1/1.5);
\draw[dualblue, thick, ]  (-1/2/1.5,1/4/1.5) --  (0 ,1/4/1.5+0.577/2/1.5);
\draw[dualblue, thick, ]  (1/2/1.5,1/4/1.5) --  (0 ,1/4/1.5+0.577/2/1.5);
\draw[dualblue, thick, ]  (-1/2/1.5,-1/4/1.5) --  (-1/2/1.5,1/4/1.5);
\draw[dualblue, thick, ]  (1/2/1.5,-1/4/1.5) --  (1/2/1.5, 1/4/1.5);
\draw[dualblue, thick, ]  (-1/2/1.5,-1/4/1.5) --  (0 ,-1/4/1.5-0.577/2/1.5);
\draw[dualblue, thick, ]  (1/2/1.5,-1/4/1.5) --  (0 ,-1/4/1.5-0.577/2/1.5);
\draw[dualblue, thick, ]  (-1/2/1.5,1/4/1.5) --  (-1/2/1.5-1/4/1.5 ,1/4/1.5+0.577/4/1.5);
\draw[dualblue, thick, ]  (1/2/1.5,1/4/1.5) --  (1/2/1.5+1/4/1.5 ,1/4/1.5+0.577/4/1.5);
\draw[dualblue, thick, ]  (-1/2/1.5,-1/4/1.5) --  (-1/2/1.5-1/4/1.5 ,-1/4/1.5-0.577/4/1.5);
\draw[dualblue, thick, ]  (1/2/1.5,-1/4/1.5) --  (1/2/1.5+1/4/1.5 ,-1/4/1.5-0.577/4/1.5);
\draw[dualblue, thick, ]   (0 ,1/4/1.5+0.577/2/1.5)--(0 ,1/4/1.5+0.577/2/1.5+1/4/1.5);
\draw[dualblue, thick, ]   (0 ,-1/4/1.5-0.577/2/1.5)--(0 ,-1/4/1.5-0.577/2/1.5-1/4/1.5);
\draw (-0.95/1.5,0.463/1.5) node {$a$};
\draw (-0.95/1.5,-0.463/1.5) node {$f$};
\draw (0.95/1.5,0.463/1.5) node {$c$};
\draw (0.95/1.5,-0.463/1.5) node {$d$};
\draw (0,1.05/1.5) node {$b$};
\draw (0,-0.95/1.5) node {$e$};
\draw (-0.75/2/1.5,1.266/2/1.5) node {$g'$};
\draw (0.75/2/1.5,1.266/2/1.5) node {$h'$};
\draw (1.4/2/1.5, 0) node {$i'$};
\draw (-1.3/2/1.5, 0) node {$l'$};
\draw (-0.75/2/1.5,-1.266/2/1.5) node {$k'$};
\draw (0.75/2/1.5,-1.266/2/1.5) node {$j'$};
\draw[thick]   (1.1/1.5, -1/1.5) -- (1.3/1.5, 0);
\draw[thick]   (1.1/1.5,  1/1.5) -- (1.3/1.5, 0);
\end{tikzpicture},
\nonumber \\
\non B^{s, g'h'i'j'k'l'}_{p, ghijkl,abcdef} &=F^{bgh}_{sh'g'} F^{chi}_{si'h'} F^{dij}_{sj'i'} F^{ejk}_{sk'j'}F^{fkl}_{sl'k'}F^{alg}_{sg'l'}.
\end{align}
The plaquette operator consists of $F$-symbols, $F^{abc}_{def}$. The $F$-symbols and the branching rules
together define the code. The $F$ symbols also define
a controlled-unitary operation; the external $a,b,c,d$ legs are the control qubits that determine the
resulting unitary $F^{abc}_d$, with matrix elements $[F^{abc}_d]_{ef}$.

In the Fibonacci code, the only non-trivial $F$-matrix is:
\be\label{F-matrix}
F^{111}_{1}=\begin{pmatrix}
\phi^{-1} & \phi^{-\frac{1}{2}}  \\
\phi^{-\frac{1}{2}}&  -\phi^{-1}
\end{pmatrix}.
\ee
All other $F$-symbols are either $1$ or $0$, depending on whether they are consistent with the branching rules [Eq.~\eqref{branching_Fibonacci} and \eqref{FmoveRel1}].
A quantum circuit implementing the $F$-operations in the Fibonacci code is shown in Fig.~\ref{fig:F-move_definition}(b) \cite{Bonesteel:2012fl}.
The circuit inside the dashed box, consisting of a 5-qubit Toffoli gate sandwiched by
 two single-qubit rotations, implements the $F$-matrix in Eq.~\eqref{F-matrix}. Here, $R_y(\pm \theta)=e^{\pm i \theta \sigma_y/2}$
are single-qubit rotations about the y-axis with angle $\theta$$=$$\tan^{-1} (\phi^{-\frac{1}{2}})$. All the other maps are taken care of by the rest of the quantum circuit.  The Fibonacci code can be implemented by repeated measurements of the vertex and
plaquette operators $Q_v$ and $B_p$ \cite{Bonesteel:2012fl, feng2015non}.

\textit{Local geometry deformation.}---The wave functions in the code space on two different triangulations (dual trivalent graphs)
$\Lambda$ ($\hat{\Lambda}$) and $\Lambda'$ ($\hat{\Lambda}'$) that differ locally can be related
by moves known as 2-2 Pachner moves (also called F-moves) and 1-3 Pachner moves,
with the following relations:
\begin{align}
\label{FmoveRel1}
&\Psi_{\hat{\Lambda}
}\left(~
\begin{tikzpicture}[baseline={([yshift=-.5ex]current  bounding  box.center)}]
\draw[dualblue,thick,] (0,0) -- (0.4,0);
\draw[dualblue,thick,] (-0.15,0.25) -- (0,0);
\draw[dualblue,thick,] (-0.15,-0.25) -- (0,0);
\draw[dualblue,thick,] (0.55,0.25) -- (0.4,0);
\draw[dualblue,thick,] (0.4,0) -- (0.55,-0.25);
\draw (0.2,0.105) node {$e$};
\draw (-0.175,0.375) node {$b$};
\draw (-0.175,-0.375) node {$a$};
\draw (0.575,0.375) node {$c$};
\draw (0.575,-0.375) node {$d$};
\end{tikzpicture}
\right)
 = \sum_f F^{abc}_{def} \
\Psi_{\hat{\Lambda}'}\left(~
\begin{tikzpicture}[baseline={([yshift=-.5ex]current  bounding  box.center)}]
\draw[dualblue,thick] (0,0) -- (0,0.8/2);
\draw[dualblue,thick,] (0,0) -- (0.5/2,-0.3/2) ;
\draw[dualblue,thick,] (-0.5/2,-0.3/2) -- (0,0);
\draw[dualblue,thick,] (0.5/2, 1.1/2) -- (0,0.8/2) ;
\draw[dualblue,thick,] (-0.5/2,1.1/2) -- (0,0.8/2);
\draw (0.24/2,0.4/2) node {$f$};
\draw (-0.8/2,1.25/2) node {$b$};
\draw (-0.8/2,-0.45/2) node {$a$};
\draw ( 0.8/2,1.25/2) node {$c$};
\draw (0.8/2,-0.45/2) node {$d$};
\end{tikzpicture}
\right)\\
\label{FmoveRel2}
&\Psi_{\hat{\Lambda}}\left(
\begin{tikzpicture}[scale=0.7, baseline={([yshift=-.5ex]current  bounding  box.center)}]
\draw[dualblue,thick,] (0,-2/3.4) -- (0,-2*0.4/3.4);
\draw[dualblue,thick,] (-1/2, 1/3.4 ) -- (-0.2,0.4/3.4);
\draw[dualblue,thick] (1/2,1/3.4) -- (0.4/2,0.4/3.4);
\draw[dualblue,thick,] (0.4/2,0.4/3.4) -- (-0.4/2,0.4/3.4);
\draw[dualblue,thick,] (-0.4/2,0.4/3.4) -- (0,-0.8/3.4);
\draw[dualblue,thick] (0.4/2,0.4/3.4) -- (0,-0.8/3.4) ;
\draw (-1.35/2,0.95/2) node {$b$};
\draw (0,-1.6/2) node {$a$};
\draw (1.35/2,0.95/2) node {$c$};
\draw (0, 0.75/2) node {$d$};
\draw (-0.56/2,-0.12/2) node {$e$};
\draw (0.56/2,-0.12/2) node {$f$};
\end{tikzpicture}
\right)
= [F^{abd}_{fce}]^* \sqrt{\frac{d_d d_f}{d_c}} \
\Psi_{\hat{\Lambda}'}\left(
\begin{tikzpicture}[scale=0.4, baseline={([yshift=-.5ex]current  bounding  box.center)}]
\draw[dualblue,thick,] (0,-4/3.4) -- (0,0);
\draw[dualblue,thick,] (-1, 2/3.4 ) -- (0,0);
\draw[dualblue,thick] (1,2/3.4) -- (0,0);
\draw (-1.25,0.8) node {$b$};
\draw (0,-1.7) node {$a$};
\draw (1.25,0.8) node {$c$};
\end{tikzpicture}\right).
\end{align}
\vspace{-0.2in}

\begin{figure}
\centering
\includegraphics[width=1\columnwidth]{braiding_intuition.pdf}
  \caption{Understanding the essence of braiding via the equivalence of local entanglement renormalization and manifold stretching/squeezing. (a) MERA is equivalent to stretching the manifold in $\text{log}_2(d)$ steps. (b) Local entanglement renormaliza- tion is equivalent to stretching the manifold in a single step. (c) Locally squeeze the rest of the region in a single step to preserve the overall shape of the manifold.  }
\label{fig:braiding_intuition}
\end{figure}

\nin Here $[F^{abd}_{fce}]^*$ denotes the complex conjugate of the elements of the $F$-move  defined in Eq.~\eqref{FmoveRel1}, while one has $[F^{abd}_{fce}]^*=F^{abd}_{fce}$ for the Fibonacci code. The local geometry deformation of the triangulation and  dual trivalent graph corresponding to the two types of Pachner moves are illustrated in Fig.~\ref{fig:F-move_definition}(c, d). Note that the F-move preserves the number of qubits (and also their locations) and is obviously a unitary transformation which can be implemented by the conditional circuit in Fig.~\ref{fig:F-move_definition}(b). On the other hand, the 1-3 Pachner move adds (entangles) three additional ancilla qubits to the code space, as shown in Fig.~\ref{fig:F-move_definition}(d) (from left to right).  The reverse process (from right to left) removes (disentangles) three qubits from the code space.    Therefore, the 1-3 Pachner move can be  related to entanglement renormalization performing either fine-graining or coarse-graining of the lattice, which has been studied in the context of MERA (\textit{multi-scale entanglement renormalization ansatz}) of string-nets \cite{Vidal:2007va, Konig:2009bm}.

Taking into account the additional qubits, the 1-3 Pachner move can also be implemented by a sequence of unitary gates as illustrated in Fig.~\ref{fig:F-move_definition}(e).
We first consider three extra qubits, each initialized
to the $|0\rangle$ state. Next, we apply a CNOT (indicated by the purple arrow), which takes $|b\rangle|0\rangle \mapsto |b\rangle |b\rangle$. This is equivalent to an isometry in the MERA language. At the same time, we apply modular $S: |0\rangle $$\mapsto $$ \sum_i \frac{d_i}{D} |i \rangle$
to the top-most qubit \footnote{In the context of Fibonacci code, the modular S gate is a single qubit rotation which can be represented by 
$S=\frac{1}{\sqrt{2+\phi}}\left(^{1, \ \ \phi}_{\phi, -1}  \right)
.$
}, which effectively builds a `tadpole diagram' connected to the original graph through the edge with the remaining
ancilla in the $\ket{0}$ state.  Note that the original edge labeled by $b$ is split into two edges with the same label $b$.
Next, we apply two successive F-moves and hence end up with the desired trivalent graph with a triangular plaquette replacing the original vertex in the center. This process is reversible.

\textit{Moving anyons in constant time.}---An intuitive way to understand the moving protocol is through the picture of \textit{local entanglement renormalization}.
The essence of entanglement renormalization and the MERA circuit can be understood as a global coarse-graining
(fine-graining) process that `merges' (`splits') several qubits together, effectively
removing (adding) qubits in the code, as illustrated in Fig.~\ref{fig:braiding_intuition}(a). In the context
of topological order, one can think of this process as squeezing (stretching) the manifold which supports
the topological states. Now one can consider anyons or defects as punctures (yellow circles) in the
manifold as illustrated in the lower panel. In order to separate two adjacent punctures to distance $d$,
one needs a MERA circuit with depth (layers) $\log_2(d)$, where each step stretches the manifold by a
factor of 2.
When the two punctures are already separated by a distance $d$, one can perform one layer of the
local entanglement renormalization circuit (with constant depth) to stretch (fine-grain) the region between the two punctures to increase the distance to $2d$,
which effectively adds qubits into the system, as illustrated in Fig.~\ref{fig:braiding_intuition}(b).
Now the manifold is effectively enlarged due to the addition of qubits.
In order to preserve the shape of the manifold away from the region of the punctures, one can also perform one layer of local entanglement renormalization to squeeze (coarse-grain) the region on the left and right sides of the punctures, as shown in Fig.~\ref{fig:braiding_intuition}(c).  Thus one effectively ends up with the same overall shape of the manifold, with
the two punctures being separated by a factor of 2, i.e., $d\rightarrow 2d$.
Note that according to the left panel of Fig.~\ref{fig:braiding_intuition}(c), in order to map the qubit lattice exactly
to the original shape, one performs SWAPs (green arrows indicated in the bottom layer) with largest distance of
$\mathcal{O}(d)$.  The long-range SWAP ensures that the actual location of each puncture is moved by a distance $d/2$.

\begin{figure}
\centering
\includegraphics[width=0.9\columnwidth]{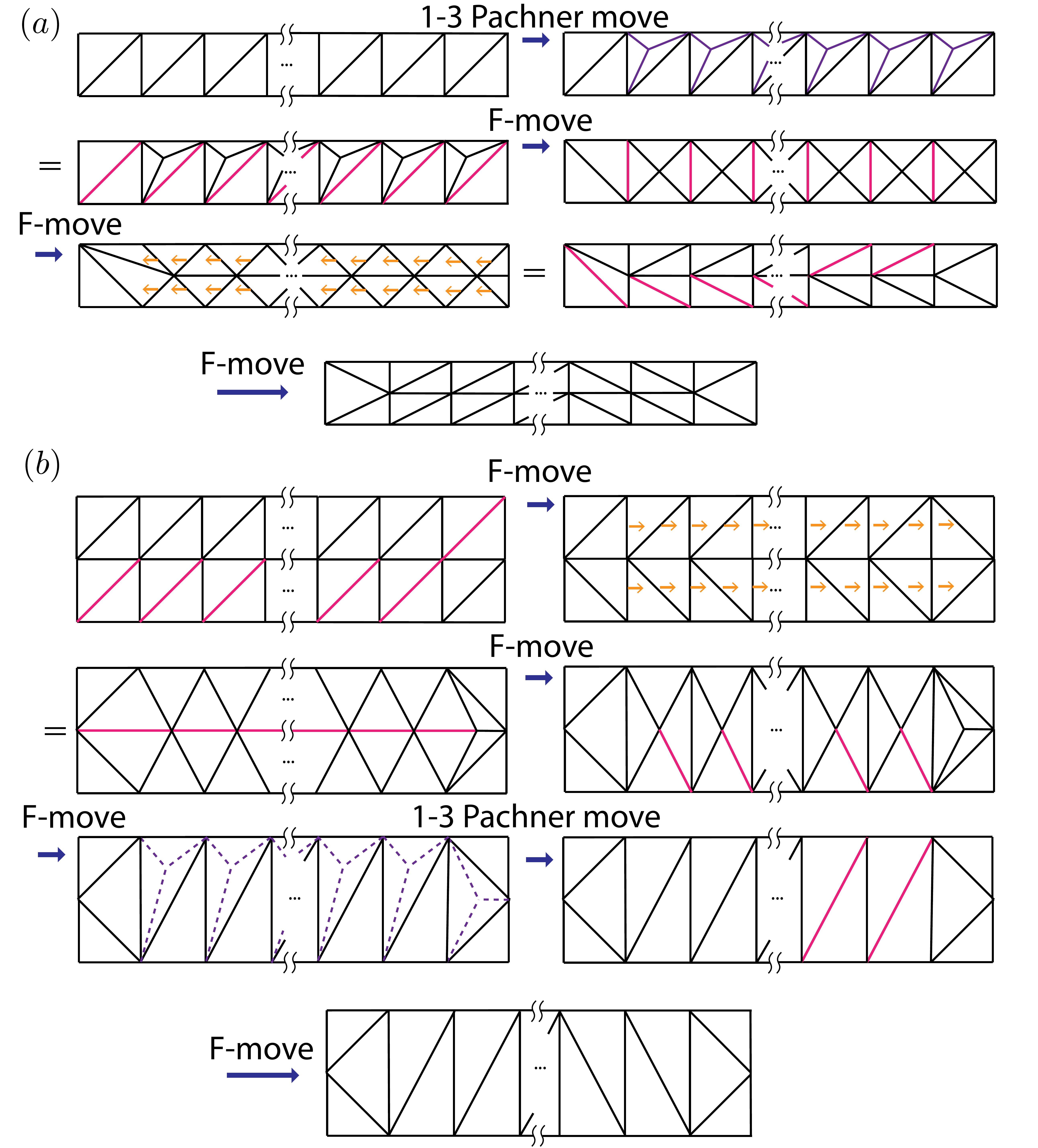}
  \caption{ Gadgets for local geometry deformation in Turaev-Viro codes.  The solid (dashed) purple lines represent added (removed) edges during the 1-3 Pachner moves.  
  The yellow arrow indicates the equivalence between two triangulations.  (a) Fine-grain the lattice via splitting a single row into two in a bounded region. (b) Coarse-grain the lattice via merging two rows into one.
\label{fig:longGadgets}
}
\end{figure}

\textit{Constant-depth braiding circuit.}---For the implementation of  braiding, we need to introduce two  elementary gadgets,
as illustrated in Fig.~\ref{fig:longGadgets}. In Fig.~\ref{fig:longGadgets}(a), we consider triangulation of a single
row of arbitrary length. By utilizing ancilla qubits, we can implement the 1-3 Pachner moves, which increase the number of vertices
of the triangulation. By a finite sequence of F-moves and local SWAPs, we can
effectively split a single row of arbitrary length $L$ into two rows, with a constant (independent of $L$)
number of steps (i.e., a constant depth local unitary circuit). In Fig.~\ref{fig:longGadgets}(b), we illustrate how
two rows can be converted into a single row by a finite number of steps.  Note that in both protocols,
the qubits on the outer boundary of the rows shown are completely unaffected, acting as control qubits
for the unitary operations, allowing the transformations to be applied to all rows in parallel.

Using the above gadgets, we can now demonstrate our braiding circuit on a triangulated region (Fig.~\ref{fig:braiding_string_net}). First, in the region between  anyon I and III, we split rows
of varying lengths in two rows, while combining rows in the region above the anyon [Fig.~\ref{fig:braiding_string_net}(b)].
We create a lattice $\Lambda'$ with a shearing pattern on the left and right sides
of anyon I; the regions above anyon I being coarse grained (squeezed) while the region below it
is fine grained (stretched). Now via long-range permutation of qubits (green arrows) $\mathcal{P}_\sigma$, where $\mathcal{P}_\sigma$
is the unitary representation of the permutation $\sigma$, we reach the configuration in Fig.~\ref{fig:braiding_string_net}(c)
which is isomorphic to the configuration in Fig.~\ref{fig:braiding_string_net}(b),
with anyon I being moved up in space.  To recover the original triangulation $\Lambda$, we apply another
retriangulation in the strip on the right of anyon I (pink thick lines), thus mapping back
to the original lattice in Fig.~\ref{fig:braiding_string_net}(d).

The above protocol, using a constant-depth local quantum circuit and long-range qubit permutations,
effectively moves one anyon vertically by a distance of the order of the separation between the nearest
anyon II (on the order of the code distance $d$). The (vertical) separation between anyon I and III
is also increased by a factor of 2, which concretely demonstrates the local entanglement renormalization idea in
Fig.~\ref{fig:braiding_intuition}(c). To complete a braiding cycle, we apply another 5 shots
of a similar procedure, which then effectively braid anyons I and II around each other as illustrated
in Fig.~\ref{fig:braiding_string_net}(e, f).  Here, we show the qubits (black dots) and trivalent graph
(light blue lines) explicitly for concreteness.

The permutations can be applied through a constant depth circuit in a variety of ways. For example, arbitrary permutations
can always be implemented by a depth-two circuit, where each layer corresponds to long-range SWAP operations
applied in parallel. This can be seen by noting that an arbitrary permutation of objects can be written as a product of cyclic permutations over disjoint sets.
A cyclic permutation can always be performed in two steps, where each step corresponds to SWAP operations applied in parallel.

To summarize, a single braiding operation can be performed in a constant number of steps, independent of the system size
and code distance. This is in contrast to previous computation schemes for the Turaev-Viro code
in Ref.~\cite{Koenig:2010do}, where braiding or Dehn twists are implemented by sequential F-moves with circuit depth of $\mathcal{O}(d)$.
Here we have demonstrated a 6-step procedure:
\vspace{-0.1in}
\begin{align}
\mathcal{B}_{\text{I},\text{II}}=\prod_{i=1}^{6} \mathcal{LU}'_i \mathcal{P}_{\sigma, i} \mathcal{LU}_i.
\end{align}
\vspace{-0.18in}

\nin Each step is composed of a  constant-depth local quantum circuit $\mathcal{LU}_i$ corresponding to a
retriangulation of the manifold, a permutation of qubits $\mathcal{P}_{\sigma, i}$ over a distance $\mathcal{O}(d)$, and
another local circuit $\mathcal{LU}'_i$ in order to retriangulate the manifold back to the original triangulation.

\begin{figure}[t]
\centering
\includegraphics[width=1\columnwidth]{braiding_string_net.pdf}
  \caption{(a-f) Braiding of two non-abelian anyons in Turaev-Viro codes with constant-depth
    circuit. The orange dashed lines in (b) and (c) show the equivalent edges before/after the permutation. Each green arrow in (e) corresponds to an instantaneous moving (number specifying the order)  carried out by an analogous procedure described in panels (a-d). }
  \label{fig:braiding_string_net}
\end{figure}

We emphasize that our protocols are constant depth unitary circuits, which neither depend on the results
of any measurement outcomes nor introduce plaquette or vertex operators whose measurement outcomes are unknown. This makes our protocols fundamentally distinct from other proposed methods for logical operations in stabilizer codes,
such as lattice surgery methods or certain schemes for moving non-Abelian defects in Abelian stabilizer codes, e.g.~in Ref.~\cite{fowler2012}. Therefore, no classical communication or classical computation is required and thus our protocols are truly constant depth both in terms of quantum operations and classical computation. 

\textit{Topological protection and fault tolerance aspects.}---The constant-depth logical gates (denoted by $U$) developed here are
naturally topologically protected (and thus can be made fault tolerant) since a local operator $O$ with support in a region $\mathcal{R}$ is mapped
 to another local operator $U^\dag O U$ supported in a region $\mathcal{R}'$ such that the area ratio of $\mathcal{R}'$ and
$\mathcal{R}$ is bounded by an $\mathcal{O}(1)$ constant factor $c$,
similar to the property of a locality-preserving
unitary \cite{Beverland:2016bi}\footnote{The difference with the definition of the locality-preserving unitary used in
  Ref.~\cite{Beverland:2016bi} is that in our case $\mathcal{R}$ and $\mathcal{R'}$ do not have to be located near each other.}, i.e.,
$\textsf{supp}(U^\dag O U) $$\le$$ c \  \textsf{supp}(O)$. This is due to the constant depth of the local unitary circuits and the specific form of the permutations we used. As a result, the circuit only changes the length of error strings by an $\mathcal{O}(1)$
constant factor independent of code distance $d$. Therefore, any local error string with length much less than $d$ remains local during the protocol and can not access the non-locally encoded logical information \footnote{See Appendix A.}.

After the application of each logical gate, the extra time overhead due to decoding and error correction depends on the details of the logical circuit of the target quantum algorithm. In the cases when the length of the error string does not increase or increases linearly, the situation is analogous to a constant-depth local quantum circuit \cite{Bravyi:2013dx} and only $\mathcal{O}(1)$ rounds of syndrome measurement per logical gate  is expected, leading to a constant time overhead. In the worst-case scenario, when certain sequence of braids are repetitively applied in the same region and the error string is always stretched by a constant factor, the error string will grow to the code distance in $\log d$ time steps. To prevent this, $\mathcal{O}(d)$ rounds of measurement needs to be inserted for every $\log d$ logical gates in the presence of measurement noise to decode the error syndrome. Therefore, we  estimate that there is an $\mathcal{O}(d/\log d)$ time overhead in the generic case to achieve full fault tolerance, although a detailed decoding scheme needs to be further developed to verify this estimation.  Due to the strictly bounded operator support of our  logical gates (Fig.~\ref{fig:braiding_string_net}), one also expects that certain highly sequential logical circuits can also have $\mathcal{O}(1)$ time overhead, when any given physical qubit is only acted on by $\mathcal{O}(1)$ operators in $\mathcal{O}(d)$ time steps (such that $d$ rounds of measurement can be applied to decode the error syndrome before the error string is further stretched).

\nin \textit{Acknowledgement.}---We thank M. Freedman, J. Preskill, M. Hastings, J. Haah,  M. Hafezi,  Z. Wang, and
S. Jordan for helpful discussions.  This work was  supported by
NSF CAREER (DMR-1753240), JQI-PFC-UMD, ARO-MURI and YIP-ONR.

\begin{appendix}

\section{Discussion of topological protection}
Here we demonstrate two key features of our braiding circuits that imply their suitability for topological protection:
(1) Any pre-existing error strings grow/shrink by at most a constant factor that is independent of code distance.
(2) The probability that a noisy implementation of our braiding circuit introduces error strings of size $l$
is proportional to $e^{-l/l_0}$ in the limit of large $l$. Here $l_0$ is a constant, independent of code distance,
that depends on details of the braiding circuit.

To understand the above statements, we first consider the constant depth local quantum circuits, denoted $\mathcal{LU}$.
These are strictly \textit{locality-preserving unitaries} \cite{Beverland:2016bi, Bravyi:2013dx}. They
are protected logical gates since the  `light cone' \cite{Lieb_Robinson} bounds the propagation of the pre-existing errors.
In other words, any pre-existing error string can grow by a length that is bounded by the depth of the circuit, while any
errors in the circuit itself can lead to new error strings whose length is also bounded by the depth of the circuit. Since the depth
is constant, independent of code distance, such circuits are inherently fault-tolerant.

The permutation $\mathcal{P}_\sigma$ used in our scheme is more non-trivial, and lies at the heart of why we are able
to circumvent various no-go theorems on locality-preserving unitaries \cite{Beverland:2016bi, Bravyi:2013dx}.
Neverthless, the statements (1) and (2) above continue to hold. The reason is that the permutations that appear in our circuits
are not generic types of permutations -- they belong to a specific class of permutations that we refer to as \textit{connectivity-preserving isomorphisms} (CPI).
While a CPI can permute qubits over long distances, it preserves the local connectivity of the underlying lattice structure
of the codes (and Hamiltonian). More concretely, for a pair of neighboring vertices $v_1$ and $v_2$ in the original lattice (triangulation)
and the permuted vertices $\sigma(v_1),  \sigma(v_2)$, the edge $e[v_1, v_2]$ connecting the original vertices $v_1$ and $v_2$ is
exactly transformed to the new edge connecting the new vertices, i.e.,  $\mathcal{P}_\sigma: e[v_1, v_2] \longmapsto e[\sigma(v_1), \sigma(v_2)]$,
which has length of $\mathcal{O}(1)$  and hence remains local.

\begin{figure}
 \includegraphics[width=1\columnwidth]{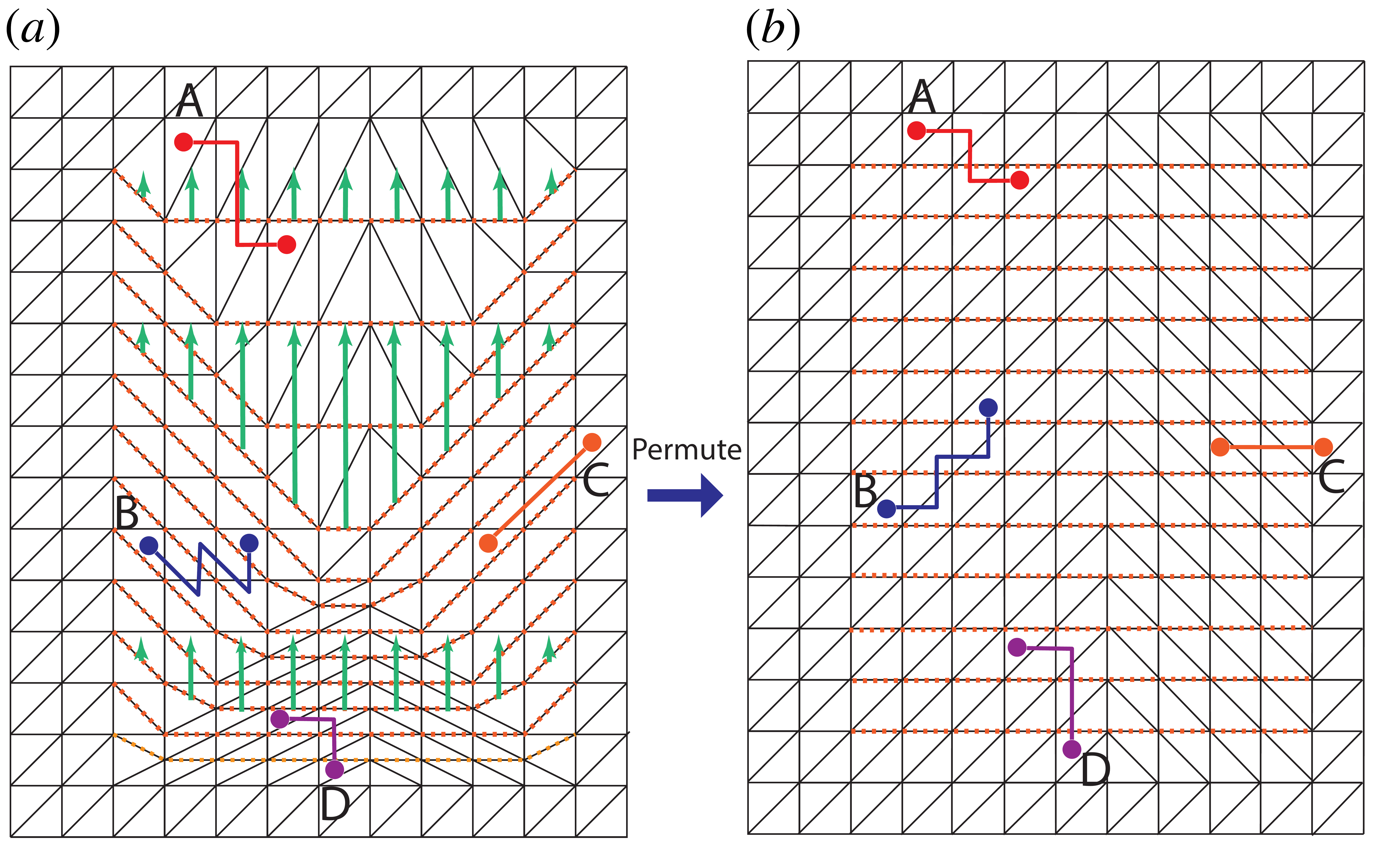}
  \caption{Error propagation due to the permutation $\mathcal{P}_{\sigma}$.}
  \label{fig:error_string_stretch}
\end{figure}

To analyze fault tolerance of a CPI, we first consider propagation of pre-existing errors under a perfect (error-free) CPI.
Let us consider an error string (the two end points of the string corresponding to anyons) with a length $l$ much smaller
than the code distance, $\textit{l} \ll d$, and which has support on sites $\{ j_1, j_2, ... , j_n \}$, as illustrated in Fig.~\ref{fig:error_string_stretch}.
Our CPI permutation $\mathcal{P}_\sigma$ maps the string onto the new sites  $\{ \sigma(j_1), \sigma(j_2), ... , \sigma(j_n) \}$ which is a deformed error string with a length $l'$. Since $\mathcal{P}_\sigma$ basically implements manifold stretching/squeezing by a constant factor (see Fig.~\ref{fig:error_string_stretch}),  $l'$ would be of the same order as before, i.e., $\mathcal{O}(\textit{l})$, which is still much less than $d$.
One can compare  the two configurations in Fig.~\ref{fig:error_string_stretch}(a,b), and see that the error string A  gets
squeezed, B and C gets deformed, and D gets stretched.  Therefore, although CPI  does not preserve the location of errors, it only
changes the length of the error string by a constant factor (independent of code distance), so that it does not introduce logical errors.  

In sum, the combined ($\mathcal{LU}$ and $\mathcal{P}_\sigma$) constant-depth logical gate (denoted by $U$)  maps a local operator $O$ with support in a region $\mathcal{R}$ to another local operator $O'=U^\dag O U$
with support in a region $\mathcal{R}'$, such that the area ratio between $\mathcal{R'}$ and $\mathcal{R}$ is bounded by an $\mathcal{O}(1)$ constant factor $c$, similar to the property of a locality-preserving unitary, namely,
\be
\textsf{supp}(U^\dag O U) \le c \  \textsf{supp}(O).
\ee
 Therefore, the constant-depth logical gate is naturally topologically protected and can be made fault-tolerant. This proves statement (1) above. The difference from the definition of the locality-preserving unitary in Ref.~\cite{Beverland:2016bi} is that here the regions $\mathcal{R}$ and $\mathcal{R}'$ do not have to be in the vicinity of each other.

To analyze statement (2), we further consider errors that can be generated when the qubit permutation has errors.
Let us consider the case where we implement the qubit permutations through a constant depth
circuit using long-range SWAP operations. As with any discussion of fault-tolerance, we have to assume a particular reasonable
noise model. Here we consider a noise model where errors in the different SWAP operations are independent of each other;
this is the analog of the uncorrelated independent noise model that is typically assumed for obtaining finite thresholds
in QECCs. Therefore, we assume that with some probability $p_{\mathcal{E}}$ our circuit implements $(1 + \mathcal{E}(i,j)) \text{SWAP}(i,j)$
instead of $\text{SWAP}(i,j)$, where $\mathcal{E}(i,j)$ is an arbitrary unitary operation that acts only on the qubits $i$ and $j$.

\begin{figure}
 \includegraphics[width=1\columnwidth]{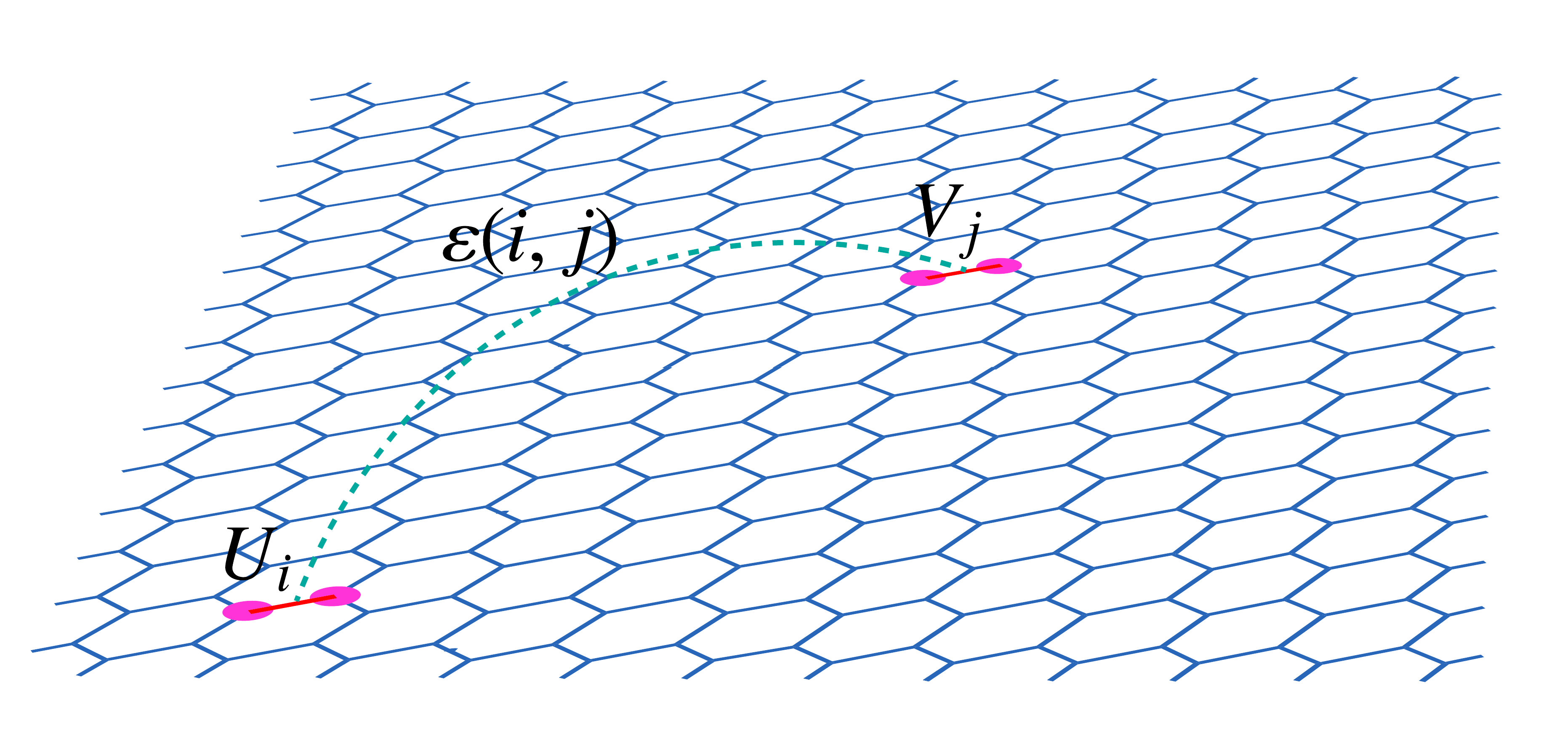}
  \caption{Illustration of two error   strings (corresponding to 4 anyonic particle excitations) with $O(1)$ length created by a geometrically nonlocal weight-two error $\epsilon(i,j)$.}
  \label{fig:illustration}
\end{figure}

Importantly, $\mathcal{E}(i,j)$ is a weight-two operator: $\mathcal{E}(i,j) = \int_{U,V} c_{i,j}(U,V) U_i \otimes V_j$,
where $U_i$, $V_i$ are single-qubit unitary operators acting on qubit $i$ and $c_{i,j}(U,V)$ are constant coefficients that define $\mathcal{E}(i,j)$.
The operator $U_i$ (respectively, $V_j$) acting on a topological state can create at most an error string of length $\mathcal{O}(1)$. Therefore,
the operator $\mathcal{E}(i,j)$ acting on a topological state creates a superposition of states, each of which has error strings
of length at most $\mathcal{O}(1)$ relative to the state before acting with $\mathcal{E}(i,j)$.
Stated differently, $(1+ \mathcal{E}(i,j))$ cannot change the topological charge of a region of
size $\mathcal{O}(1)$ surrounding the site $i$, or the topological
charge of a region of size $\mathcal{O}(1)$ surrounding the site $j$.  
A specific example shown in Fig.~\ref{fig:illustration} illustrates that the $U_i$ and $V_j$ violate the branching rules on the vertices located on both ends of edge $i$ and edge $j$ respectively.  In this situation, two error string of length $1$ corresponding to four anyons even though they are created by the geometrically non-local error $\mathcal{E}_{i,j}$.

In order to create error strings of length $l$, we need to apply a string of error operators, e.g.
$\mathcal{E}_1(i_1,j_1) \mathcal{E}_2(i_2,j_2) \cdots \mathcal{E}_\ell(i_\ell,j_\ell) $, where either $i_1, i_2, i_3,\cdots$ or $j_1, j_2, j_3, \cdots$
form a connected path of length $l$. However the probability of this occurring for any given path is $\prod_{i = 1}^l p_{\mathcal{E}_i}$. We can thus
conclude that the probability of creating an error string of length $l$ is exponentially small, $\propto e^{-l/l_0}$, for some
constant $l_0$. This proves statement (2).

The essential point of the above result is that while the SWAP operators are long-ranged, they are
\it low weight \rm (specifically weight 2) operators. However, creating error strings of length $l$ requires an operator of weight $\sim l$.
This guarantees that as long as the different SWAP operators have independent errors, the probability of creating error strings of length
$l$ is exponentially small in $l$.

\section{Fault tolerance aspects and the overall time complexity}
Although the logical gate applied here is constant-depth itself, differing from the situation of conventional braiding,  extra time overhead might be introduced due to the error correction and decoding processes after the application of the gate in order to achieve full fault tolerance.  Without measurement error,  the error syndrome can be decoded immediately after the application of the gate, since the error string is only changed by an $\mathcal{O}(1)$ constant factor independent of the code distance $d$, and hence remains correctable. 

However, in the presence of faulty measurement, one is expected to perform $\mathcal{O}(d)$ rounds of measurements to decode the error syndrome, similar to the case of surface code error correction.  In the context of transversal gates or constant-depth local quantum circuits \cite{Bravyi:2013dx}, the error string either does not change or is stretched at most linearly, and one hence expects that only $\mathcal{O}(d)$ rounds of measurements are needed for every $\mathcal{O}(d)$ times of the application of such logical gates. The overall time overhead is hence $\mathcal{O}(1)$ in these situations.  In the context of our logical gates, which are geometrically non-local, the overall time overhead depends on the detailed property of the logical circuits of the corresponding quantum algorithms.   In the cases when the error string is stretched linearly with time, including the situation of applying certain highly sequential logical circuits,  the overall time complexity is still  $\mathcal{O}(1)$. In the worst case, when certain sequence of braids are repetitively applied in the same region and stretch the error string exponentially with time which leads to it growing to the code distance $d$ in $\log d$ time steps. Therefore, one needs $\mathcal{O}(d)$ rounds of syndrome measurements for every $\mathcal{O}(\log d)$ times of application of our logical gates, giving rise to an overall $\mathcal{O}(d/ \log d)$ time overhead.  More concrete examples are discussed in Ref.~\cite{Zhu:2018CodeLong}.

\section{Experimental platforms}
A number of experimental platforms with long-range connectivity enable the
physical implementation of long-range SWAP operations in one time step (between two consecutive syndrome measurements):

\textbf{I.}~\textit{Long-range connectivity in ion traps} mediated by vibrational modes of
ions \cite{Linke:2017bz}.
\textbf{II.}~\textit{Modular architecture of 3D superconducting cavities} \cite{CampagneIbarcq:2017wq, Axline:2017uq, Chou:2018vz}, using reconfigurable long-range connectivity between cavity nodes, routed by microwave circulators and superconducting cables \cite{CampagneIbarcq:2017wq}.
One scheme is through direct quantum state transfer between remote cavity nodes in a network, equivalent to a
long-range SWAP \cite{Axline:2017uq}. The noise is uncorrelated if different cables are used for individual
SWAP processes. An alternative is through remote entanglement generation and teleportation \cite{CampagneIbarcq:2017wq, Chou:2018vz},
which also has uncorrelated noise for individual teleportation channels.
\textbf{III.} ~\textit{Circuit QED with cavity buses}. Here, long-range interaction between
superconducting qubits or semiconductor spin qubits can be mediated by cavity array serving as quantum buses \cite{Majer:2007em, Helmer:2009de, Naik:2017bo}.

Another method to implement the long-range permutations required in our circuit is to physically
move the qubits to their desired positions. For example, high-fidelity fast shuttling of individual ion qubits has been realized
experimentally  \cite{Bowler:2012fr,  Wright:2013df} and  proposed  for a scalable quantum computation
architecture \cite{Home:jr, Lekitsch:2015ua}.  The individual shuttling processes have independent noise, e.g., ion
heating \cite{Wright:2013df} (as the dominant error source). In this case, as long as the movement of a qubit
from $j \mapsto \sigma(j)$ does not induce errors along the entire path that connects $j$ to $\sigma(j)$,
but rather only induces errors that can be modeled by a low-weight operator such as $\mathcal{E}(i,j)$ discussed above,
our statement (2) will continue to hold in this case as well.

\end{appendix}

\end{document}